\documentclass[10pt,letterpaper]{article}
\usepackage[margin=1in]{geometry}
\usepackage{ifpdf}
\ifpdf
    \usepackage[pdftex]{graphicx}
    \graphicspath{{figs/}{matlab/}}   % still needed, even with Inkscape-Pdf-Latex
    \usepackage[update]{epstopdf}
\else
	\usepackage{graphicx}
	\graphicspath{{figs/}{matlab/}}   % still needed, even with Inkscape-Pdf-Latex
\fi
\usepackage{pdflscape}
\usepackage{marvosym}
\usepackage{enumitem,color}
\usepackage{amsthm}
\newtheorem{theorem}{Theorem}

\newtheorem{corollary}{Corollary}

\usepackage{hyperref}
\usepackage{array}
\usepackage{mathrsfs,bm,bbm}
\usepackage{amsmath,amsthm}
\usepackage{amsfonts}
\usepackage{amssymb}
\usepackage{cite,setspace}
\usepackage{breqn}
\usepackage{multirow}

\newcommand{\Hc}{\mathcal{H}}

\newcommand{\Wc}{\mathcal{W}}
\newcommand{\Xc}{\mathcal{X}}
\newcommand{\Yc}{\mathcal{Y}}
\newcommand{\Zc}{\mathcal{Z}}

\newcommand{\Eb}{\mathbb{E}}
\newcommand{\Rb}{\mathbb{R}}

\newcommand{\KL}{\mathrm{KL}}

\newcommand{\dm}{\mathrm{d}}

\newcommand{\gen}{\mathrm{gen}}

\DeclareMathOperator{\Tr}{Tr}

\usepackage{xparse}
\NewDocumentCommand{\ml}{m O{M} O{Q_n^{[M]}}} {\mathcal{L}_{#1}(#2\rightarrow #3)}

\title{Exactly Tight Information-Theoretic Generalization Error Bound for the Quadratic Gaussian Problem}
\author{Ruida Zhou, Chao Tian, and Tie Liu
}
\date{}

\begin{document}
\maketitle

\begin{abstract}
We provide a new information-theoretic generalization error bound that is exactly tight (i.e., matching even the constant) for the canonical quadratic Gaussian (location) problem. Most existing bounds are order-wise loose in this setting, which has raised concerns about the fundamental capability of information-theoretic bounds in reasoning the generalization behavior for machine learning. The proposed new bound adopts the individual-sample-based approach proposed by Bu et al., but also has several key new ingredients. Firstly, instead of applying the change of measure inequality on the loss function, we apply it to the generalization error function itself; secondly, the bound is derived in a conditional manner; lastly, a reference distribution is introduced. The combination of these components produces a KL-divergence-based generalization error bound. We show that although the latter two new ingredients can help make the bound exactly tight, removing them does not significantly degrade the bound, leading to an asymptotically tight mutual-information-based bound. We further consider the vector Gaussian setting, where a direct application of the proposed bound again does not lead to tight bounds except in special cases. A refined bound is then proposed for decomposable loss functions, leading to a tight bound for the vector setting.
\end{abstract}

\section{Introduction}

Understanding the generalization behavior and bounding the generalization error of learning algorithms are important subjects of study in machine learning theory. Recently, information-theoretic approaches to bound generalization errors have drawn considerable attention in both the information theory community and the machine learning community \cite{russo2016controlling, xu2017information,asadi2018chaining,pensia2018generalization, issa2019strengthened, negrea2019information, jose2021information,wu2020information,bu2020tightening,steinke2020reasoning,haghifam2020sharpened,hellstrom2020generalization,wu2022fast,zhou2022individually,rodriguez2021random,zhou2022stochastic,aminian2021exact,
barnes2022improved,aminian2022tighter,haghifam2022understanding,hafez2020conditioning,hellstrom2022evaluated,haghifam2023limitations,wang2023generalization,wang2023tighter,wang2021generalization}.
These bounds can provide intuitions by relating to information-theoretic quantities, leading to novel reasoning and revealing deep connections to existing results such as the classic VC-dimension and Rademacher complexity \cite{shalev2014understanding}. Information-theoretic bounds can take into account both data distribution and the dependence between data and algorithm output, which cannot be fully captured by the conventional complexity-based bounds. 

In classic information theory research, the study of complex communication systems usually starts from the simplest canonical settings. Particularly, the canonical quadratic Gaussian settings have played tremendous roles in the study of both channel coding and source coding \cite{CoverThomas}. The study of Gaussian channel under the average power constraint can be traced back to the original paper by Shannon \cite{Shannon:48} and led to many subsequent developments in wireless communications \cite{tse2005fundamentals}. Similarly, the Gaussian source compression under the quadratic distortion measure has been studied extensively \cite{berger2003rate,gray1989source}, which led to many well-used designs of data compression and quantization methods. The motivation to study the Gaussian settings can perhaps be explained as follows. Mathematically, the simplicity of the Gaussian settings, the statistic properties of Gaussian distributions, the optimality of linear estimators, and the connection to information measures (e.g., differential entropy and entropy power inequality) allow the derivation of precise results and exact tight bounds, which can serve as a running ramp for more complex settings. Practically, Gaussian noises and Gaussian sources can be good approximations to random quantities encountered in many applications, further strengthening the motivation to study Gaussian settings. 

In sharp contrast to the classical information theory research, in the study of generalization error bounds, although various more sophisticated settings such as meta-learning \cite{jose2021information,hellstrom2022evaluated} and iterative stochastic algorithms \cite{pensia2018generalization,rodriguez2021random} have been considered, our understanding of the canonical quadratic Gaussian setting is in fact quite limited. In this setting, independent Gaussian samples are observed, and the learning algorithm chooses the sample average as the hypothesis parameter to locate the mean value. The loss function is the squared difference between the samples and this hypothesis parameter. It turns out that earlier information-theoretic bounds are either vacuous \cite{xu2017information} or order-wise loose \cite{bu2020tightening,steinke2020reasoning,zhou2022individually,hellstrom2020generalization}. The only approaches that provide order-wise tight bounds in this setting either only hold asymptotically \cite{wu2022fast}, or have a loose constant and require a careful construction of certain auxiliary probability structure \cite{zhou2022stochastic}. 

In this work, we provide a new information-theoretic bound that is exactly tight (i.e., matching even the constant) for the canonical quadratic Gaussian (location) problem. The proposed new bound adopts the individual-sample-based approach proposed by Bu et al. \cite{bu2020tightening}, but also has several key new ingredients. Firstly, instead of applying the change of measure inequality on the loss function, we apply it to the generalization error function itself; secondly, the bound is derived in a conditional manner; lastly, a reference distribution, which bears a certain similarity to the prior distribution in the Bayesian setting, is introduced. The combination of these components produces a general  KL-divergence-based generalization error bound. We also show that although the conditional bounding and the reference distribution can make the bound exactly tight, removing them does not significantly degrade the bound, which results in a mutual-information-based bound that is also asymptotically tight in this setting. 

In order to further understand the proposed generalization error bound, we consider the vector version of the Gaussian location problem. The samples here are independent Gaussian vectors, and the algorithm is again the sample mean, but the loss function is a general squared matrix norm. We show that a direct application of the proposed bound is no longer tight in this setting except in certain special cases. However, a refined information-theoretic bound that takes advantage of the decomposability of the matrix norm can indeed lead to a tight bound. 

The rest of the paper is organized as follows. In Section \ref{sec:pre} we provide the preliminaries and some relevant previous results. The new generalization error bound is provided in Section \ref{sec:newbound}, and then applied on the canonical quadratic Gaussian setting in Section \ref{sec:Gaussian}. The generalized vector setting is considered in Section \ref{sec:vector}. Finally, Section \ref{sec:conclusion} concludes the paper, and a few technical proofs are included in the appendix. 

\section{Preliminaries}
\label{sec:pre}

\subsection{Generalization Error}
Denote the data domain as $\Zc$, e.g., in the supervised learning setting $\Zc = \Xc \times \Yc$, where $\Xc$ is the feature domain and $\Yc$ is the label set. The parametric hypothesis class is denoted as $\Hc_\Wc = \{ h_{W} : W \in \Wc \}$, where $\Wc$ is the parameter space. During training, the learning algorithm has access to a sequence of training samples  $Z_{[n]} = (Z_1, Z_2, \ldots, Z_n)$, where each $Z_i$ is drawn independently from $\Zc$ following some unknown probability distribution $\xi$. The learner can be represented by $P_{W | Z_{[n]}}$, which is a kernel (channel) that (potentially randomly) maps $\Zc^n$ to $\Wc$. 

The learner wishes to choose a hypothesis $w \in \Wc$ to minimize the following population loss, under a given loss function $\ell:  \Wc  \times \Zc \rightarrow \Rb$, 
\begin{align}
L_{\xi}(w) = \Eb_{\tilde{Z} \sim \xi}[\ell(w, \tilde{Z})]. \label{eqn:def-pop-loss} %\int_{\Zc} \ell(w, z) \xi(d z). 
\end{align}
The empirical loss of $w$ is 
\begin{align}
L_{Z_{[n]}}(w) = \frac{1}{n} \sum_{i = 1}^n \ell(w, Z_i). \label{eqn:def-emp-loss}
\end{align}
The expected generalization error of the learner $P_{W|Z_{[n]}}$ is
\begin{align}
\gen(\xi, P_{W|Z_{[n]}}) \triangleq \Eb_P \left[L_{\xi}(W) - L_{Z_{[n]}}(W) \right], \label{eqn:def-gen}
\end{align}
where the expectation is taken over the distribution $P_{W, Z_{[n]}}$ as the joint distribution implied by the kernel $P_{W | Z_{n]}}$ and the marginal $P_{Z_{[n]}} = \xi^n$.
%the joint distribution $P(W,Z_{[n]})=\xi^n P_{W|Z_{[n]}}$. 
%This quantity captures the effect of the learner's expected overfitting error due to limited training data, which we shall study in this work. 
%
%Formally we write $\Eb_{X \sim P}[f(X)] = \int_{\Xc} f(x) dP(x) $ as the expectation of $f(X)$. When the distribution of $X$ is clear from the context, we omit $P$ and write it as $\Eb_{X}[f(X)]$, where the subscript $X$ means the expectation is taken with respect to the random variable $X$. When the random variable $X$ in $f(X)$ is also clear from the context, we simply write it as $\Eb[f(X)]$.

Assume another distribution $Q_{W,Z_{[n]}}$, where $W$ and $Z_{[n]}$ are independent and the marginal $Q_{Z_{[n]}}$ is the same as $P_{Z_{[n]}}$, i.e., $Q_{W,Z_{[n]}}=Q_{W}Q_{Z_{[n]}}=Q_{W}P_{Z_{[n]}}$. The marginal distribution $Q_W$ can be viewed as a prior distribution in this case\footnote{In the Bayesian setting, the distribution $P$ is usually used to denote the prior distribution and $Q$ as the posterior (data dependent) distribution. This is reversed from ours, which follows the convention in information-theoretic literature.}. For such $Q$'s, apparently, we have
\begin{align}
\gen(\xi, Q_{W|Z_{[n]}}) \triangleq \Eb_Q \left[L_{\xi}(W) - L_{Z_{[n]}}(W) \right]=0, \label{eqn:def-gen-Q}
\end{align}
where the equality is because $Q_{W,Z_{[n]}}=Q_{W}P_{Z_{[n]}}$.

\subsection{Variational Representation of the KL Divergence}
The Donsker-Varadhan variational representation of KL divergence for a random scalar-valued random function $F=f(X)$ on a random variable $X$ is given by 
\begin{equation}
\begin{aligned}
&\KL(P || Q) = \sup_{f} \left\{\lambda \Eb_P[F] - \ln \Eb_Q[e^{\lambda F}] \right\}, \\
&\qquad\qquad \text{where equality achieved when } \lambda F^* = \ln \frac{\dm P}{\dm Q}+C, \label{eqn:variational}
\end{aligned}
\end{equation}
or in the inequality form
\begin{align}
\lambda \Eb_P[F] \leq \KL(P || Q) + \ln \Eb_Q[e^{\lambda F}],\quad \forall \lambda\in \mathbb{R}. \label{eqn:changeofmeasure}
\end{align}
This inequality is sometimes also referred to as the change of measure inequality. $P$ and $Q$ can be the distributions of the underlying random variable $X$, or more directly, the distributions of $F$. 
In the context of bounding generalization error, examples are $F = \ell(W, Z)$ or $F=L_{\xi}(W)-\ell(W,Z)$. {We remark here that in the variational representation (\ref{eqn:variational}), the supremum is taken over the functions $f$, whereas when we apply the change of measure inequality (\ref{eqn:changeofmeasure}), the function $f$ is usually already fixed, but the distribution $Q$ can be optimized to make the bound tighter. %, however, when the distribution $Q$ is constrained due to problem setting, it is not guaranteed that there exists a sequence of distributions $Q_n$'s such that the change of measure inequality can approach equality. Therefore, the change of measure inequality must be applied in a more strategic manner to yield stronger bounds. 
}

%{\color{blue}Ruida's mumbling comments (not included in the main paper): Variational inequality views $f$ as variables, however, we would like to view $Q$ as variables. Due to equation $\lambda F^* = \ln \frac{d P}{d Q}$, for any function $\lambda f$, we can find a distribution with density $\dm Q_{\lambda f} \propto \exp(- \lambda f) \dm P $. From this perspective, it even gives us a way of variational prior optimization:
%\begin{align}
%\dm Q^* \in \min_{\dm Q \in \{\text{data independent distributions} \} } \Eb_{Z}[\KL( \exp(- \lambda f) \dm P || \dm Q)].
%\end{align}
%From this perspective, it is also easy to verify the achievability of tightness: the bound can be tight, i.e., exists prior $Q$ independent of data, as long as there is a function $a(W) + b(Z)$ lies in the function class $\{ \ln \dm P(W | Z) - \lambda F(W, Z): \lambda \geq 0\}$, i.e., decoupled. We thus only need to check the coupled terms. It is not hard to verify the tightness of scalar case by this condition in (\ref{eqn:Fi})-(\ref{eqn:QW}). We can also consider vector case in (\ref{eqn:Fi-vec})-(\ref{eqn:QW-vec}). For the coupled terms of $(Z_i, W)$, $F_i = 2 W^\top A^\top Z_i + func(W) + func(Z)$, and $\log \dm P = Z_i + \alpha_i W^\top (\sum_{j\neq i} \alpha_j^2 \Sigma +\sigma_N^2\mathbf{I})^{-1}  Z_i + func(W) + func(Z)$. 
%}

The centered cumulant generating function of a random variable $F$ is
\begin{align}
\Lambda_{F, Q}(\lambda) = \ln \Eb_Q\left[e^{\lambda F}\right] - \lambda \Eb_Q[F] .
\end{align}
%If $F=f(W,Z)$, the conditional version of the centered cumulative generating function of $F$ is
Combining it with the inequality above gives 
\begin{align}
\KL(P || Q) +\Lambda_{F, Q}(\lambda)\geq \lambda \Eb_P[F]- \lambda \Eb_Q[F],\quad \lambda \in \mathbb{R}.
\end{align}
Now if we choose $F=f(W,Z)$, then for any $Z=z$ the conditional version of the above inequality is 
\begin{align}
&\KL(P_{W|Z=z} || Q_{W|Z=z}) +\Lambda_{F|Z=z, Q_{W|Z=z}}(\lambda) \geq \lambda \Eb_P[F|Z=z]- \lambda \Eb_Q[F|Z=z],\quad \lambda \in \mathbb{R},\label{eqn:conditionalDV}
\end{align}
where
\begin{equation}
\begin{aligned}
&\Lambda_{F|Z=z, Q_{W|Z=z}}(\lambda)=\ln \Eb_{Q_{W|Z=z}}\left[e^{\lambda F}|Z=z\right] - \lambda \Eb_{Q_{W|Z=z}}[F|Z=z].
\end{aligned}
\end{equation}
We will simply replace $Z=z$ in the condition with $Z$ when the exact conditional value realization is not specified. 

%With a negative $\lambda$ we therefore obtain
%\begin{align}
% \Eb_Q[F] - \Eb_P[F] &\leq \inf_{\lambda < 0} \left\{\frac{\KL(P||Q) + \Lambda_{F, Q}(\lambda) }{-\lambda} \right\} \notag\\
% &=  \inf_{\lambda > 0} \left\{\frac{\KL(P||Q) + \Lambda_{-F, Q}(\lambda) }{\lambda} \right\}, \label{eqn:bounda2}
%\end{align}
%where equality is achieved if and only if
%\begin{align}
%\ln \frac{\dm P}{\dm Q} \in \left\{ \lambda F + b : \lambda \in \Rb_-,~ b \in \Rb \right\}. \label{eqn:equalitycondition}
%\end{align}
%The equality condition can also be interpreted as requiring us to choose $\dm Q \propto \exp(- \lambda F) \dm P$.
%Similarly, 
With a positive $\lambda$, we obtain
\begin{align}
 \Eb_P[F] - \Eb_Q[F] &\leq \inf_{\lambda > 0} \left\{\frac{\KL(P||Q) + \Lambda_{F, Q}(\lambda) }{\lambda} \right\}, \label{eqn:bounda1a}
\end{align}
where equality is achieved if and only if
\begin{align}
\ln \frac{\dm P}{\dm Q} \in \left\{ \lambda F + b : \lambda \in \Rb_+,~ b \in \Rb \right\}. \label{eqn:equalityconditiona}
\end{align}
The equality condition can also be interpreted as requiring us to choose $\dm Q \propto \exp(- \lambda F) \dm P$. When $P$ is the joint distribution of underlying random variables, and $Q$ is the product distribution of their marginals, then $\KL(P||Q)$ reduces to a mutual information term.

To be consistent with past results in the literature, we will sometimes use the following definition. The Legendre dual function on the interval $[0,b)$ for some $0<b\leq \infty$ is 
\begin{align}
\Lambda^*(x) \triangleq \sup_{\lambda \in [0,b)}(\lambda x-\Lambda(\lambda)).
\end{align}
$\Lambda(\lambda)$ is convex and $\Lambda(0)=\Lambda'(0)=0$. It can be shown that the inverse dual function is
\begin{align}
\Lambda^{*-1}(y)=\inf_{\lambda\in[0,b)}\left(\frac{y+\Lambda(\lambda)}{\lambda}\right).
\end{align}

\subsection{The Scalar Quadratic Gaussian Location Problem}

In the canonical Gaussian location problem introduced by Bu et al. \cite{bu2020tightening}, data samples are $Z_1, Z_2, \ldots, Z_n \overset{i.i.d.}{\sim} \xi = 
\mathcal{N}(\mu, \sigma^2)$ and the sample-average algorithm chooses the following hypothesis $W = \frac{1}{n} \sum_{i = 1}^n Z_i$. The loss function is the quadratic function given as $\ell(w,z_i)=(w-z_i)^2$. Then the expected generalization error is
\begin{align}
\gen(\xi, P_{W|Z_{[n]}})  &= \Eb\left[(\tilde{Z}-W)^2- \frac{1}{n} \sum_{i=1}^n(Z_i - W)^2 \right]\notag\\
&=\Eb\left[ \frac{1}{n} \sum_{i=1}^n \left[  (\tilde{Z}_i-W)^2-(Z_i - W)^2 \right]\right]\notag\\
&=\frac{1}{n} \sum_{i=1}^n \Eb \left[  \tilde{Z}^2_i-Z_i^2+2(Z_i-\tilde{Z}_i)W\right]\notag\\
&= \frac{1}{n} \sum_{i=1}^n \Eb\left( \sigma^2 + \mu^2 -Z_i^2  + 2(Z_i - \mu) W \right),
\end{align}
where $\tilde{Z}_{[n]}$ are $n$ i.i.d. testing samples, independent of everything else, and the expectation is with respect to distribution $P_{\tilde{Z}} P_{Z^n,W}$, where the joint distribution $P_{Z^n,W}$ is induced by the algorithm $W = \frac{1}{n} \sum_{i = 1}^n Z_i$. It is straightforward to show that the true generalization error is in fact $2\sigma^2/n$. 

In this work, we shall in fact consider a slightly more general version of the sample-average algorithm that $W=  \sum_{i = 1}^n \alpha_i Z_i+N$, where $N$ is a Gaussian noise $\sim \mathcal{N}(0,\sigma_N^2)$, independent of $Z_{[n]}$, and $\alpha_i$'s are nonnegative weights such that $\sum_{i = 1}^n \alpha_i =1$. It can be shown that the true generalization error is also $2\sigma^2/n$ (see the Appendix). 

\subsection{Existing Generalization Error Bounds}

Xu and Raginsky, motivated by a previous work by Russo and Zou \cite{russo2016controlling}, provided a mutual information (MI) based bound on the expected generalization error \cite{xu2017information}. 
%\begin{theorem}[MI Bound \cite{xu2017information}] \label{thm:MI} 
Assuming $\ell(w, Z)$ is $\sigma$-sub-Gaussian\footnote{We call a distribution $\sigma$-sub-Gaussian if it has a variance proxy of $\sigma^2$.}  under $\xi$ for all $w \in \Wc$, then the bound is 
\begin{align}
\gen( \xi, P_{W|Z_{[n]}} ) \leq \sqrt{\frac{2 \sigma^2}{n} I\left(W; Z_{[n]} \right)}. \label{eqn:MI}
\end{align}
%\end{theorem}
One issue with this bound is that it can be vacuous, i.e., the mutual information term can be bounded. Indeed, for the quadratic Gaussian case, it is vacuous when $N=0$. Bu et al. \cite{bu2020tightening} noticed that 
the generalization error can be written as 
\begin{align}
\gen( \xi, P_{W | Z_{[n]}} ) &= \frac{1}{n} \sum_{i = 1}^n \Eb\left[ (\ell(W, \tilde{Z}_i) - \ell(W, Z_i)) \right] \label{eqn:gen-ind}\\
&=\frac{1}{n} \sum_{i = 1}^n \Eb\left[ L_{\xi}(W) - \ell(W, Z_i)) \right],\label{eqn:gen-ind1}
\end{align}
where $\tilde{Z}_i$ are independent testing data samples that are independent of $W$. The following bound can then be obtained by bounding each summand 
\begin{align}
\gen( \xi, P_{W|Z_{[n]}} ) \leq \frac{1}{n} \sum_{i = 1}^n \sqrt{2\sigma^2 I\left(W; Z_i \right)},
\end{align}
assuming $\ell(\tilde{W},\tilde{Z})$ is $\sigma$-sub-Gaussian, where $\tilde{W}$ and $\tilde{Z}$ are independent but have the same marginal distribution as that in $P_{W,Z_{[n]}}$. This bound improves upon the bound in \cite{xu2017information}, and it is in general not vacuous. However, for the quadratic Gaussian setting, it leads to an order $\mathcal{O}(1/\sqrt{n})$ bound, which is order-wise loose.
%\begin{theorem}[IMI Bound \cite{bu2020tightening}] \label{thm:IMI} 
%Suppose $\psi_{-}$ is an upper bound of $\psi_{-\ell(\tilde{W}, \tilde{Z}_i)}$, then
%\begin{align}
%\gen( \xi, P_{W|Z_{[n]}} ) \leq \frac{1}{n} \sum_{i = 1}^n \psi^{*-1}_{-}\left(I\left(W; Z_i \right) \right),
%\end{align}
%where $\tilde{W}$ and $\tilde{Z}_i$ are independent random variables that have the same marginal distributions as $W$ and $Z_i$, respectively.
%\end{theorem}

Steinke and Zakynthinou \cite{steinke2020reasoning} introduced a conditional-mutual-information-based generalization error bound. We will not provide the precise bound here, but it can be shown straightforwardly that their bound leads to an order $\mathcal{O}(1)$ bound, which is order-wise loose. Different improvements on this conditional mutual information bound have been proposed \cite{haghifam2020sharpened,rodriguez2021random,zhou2022individually}, however, in the quadratic Gaussian setting, they led to either $\mathcal{O}(1)$ or $\mathcal{O}(1/\sqrt{n})$ bounds, thus also order-wise loose. Details can be found in \cite{zhou2022individually}.

Zhou et al.  \cite{zhou2022stochastic} proposed a chaining technique to tighten the generalization error bound, and showed that with a specially constructed chain in the quadratic Gaussian setting, the bound in \cite{bu2020tightening} can be tightened to the order $\mathcal{O}(1/n)$, but with a loose constant factor. In a more recent work \cite{wu2022fast}, Wu et al. proposed a new bound assuming the function $r(\tilde{W},\tilde{Z})=\ell(\tilde{W},\tilde{Z})-\ell(w^*,\tilde{Z})$ is $\sigma^2$-sub-Gaussian, where $w^*$ is the optimal solution of the true risk. For the quadratic Gaussian setting, this bound is asymptotically optimal\footnote{The bound is only asymptotically optimal, (in fact, only asymptotically valid) since one of the inequalities is replaced by an approximation that only holds in an asymptotic manner to yield the bound. Strictly speaking, their bound can be stated as follows: for any $\epsilon>0$, for sufficiently large $n$,  the generalization error $\leq 2(1+\epsilon)\sigma^2/n$ in this quadratic Gaussian setting.}, but not optimal for finite $n$. Moreover, the function $r(\tilde{W},\tilde{Z})$ relies on the optimal solution $w^*$. A more detailed summary of the quadratic Gaussian location setting can be found in \cite{hellstrom2023generalization}.

%, which can be problematic by itself. 
% is 
%\begin{align}
%\frac{\sigma^2}{n}\sum_{i=1}^n\sqrt{\frac{4}{n}\log\frac{n}{n-1}},
%\end{align}
%which is 
%again has the optimal order $\mathcal{O}(1/n)$, but loose in the constant factor. 

% In their approach, $Z_{[n]}^\pm \triangleq (Z^{\pm1}_{1}, Z^{\pm1}_{2}, \ldots, Z^{\pm1}_{n})$ is a $2 \times n$ table of samples that each $Z_{i}^{s}$, for $s = -1,1$ and $i = 1,\ldots, n$ is independently drawn following $\xi$. The training vector $(Z^{R_1}_{1}, Z^{R_2}_{2}, \ldots, Z^{R_n}_{n})$ is selected from the table $Z^{\pm}_{[n]}$, where $R_i$'s are independent Rademacher random variables, i.e., $R_i$ takes $1$ or $-1$ equally likely. The vector $R_{[n]} = (R_1, \ldots, R_n) \in \{-1, 1\}^n$ essentially selects one sample from each column in the table, which partitions $Z_{[n]}^{\pm}$ into a training vector and a testing vector. For simplicity, we shall write $Z_i^{-1}$ and $Z_{i}^{+1}$ as $Z_i^{-}$ and $Z_{i}^{+}$, when the meaning is clear from the context. 

\section{A New Information-Theoretic Generalization Error Bound}
\label{sec:newbound}

The new information-theoretic generalization error bound is summarized in the following theorem. 
\begin{theorem}
\label{theorem:main}
Let $F_i=L_{\xi}(W) - \ell(W,Z_i)$, then we have
\begin{align}
\gen(\xi, P_{W|Z_{[n]}}) 
&\leq\frac{1}{n}\sum_{i=1}^n \Eb_{P_{Z_i}}\left[\inf_{\lambda > 0} \frac{\KL(P_{W|Z_i}\|Q^i_{W})+\Lambda_{F_i|Z_i,Q^i_{W}}(\lambda)}{\lambda}\right]\notag\\
&=  \frac{1}{n}\sum_{i=1}^n \Eb_{P_{Z_i}}\left[\Lambda^{*-1}_{F_i|Z_i,Q^i_{W}}\left(\KL(P_{W|Z_i}\|Q^i_{W})\right)\right],
\end{align}
for any $Q^i_{W,Z_i}=Q^i_WP_{Z_i}$, $i=1,2,\ldots,n$, i.e., a distribution $Q^i$ where $W$ is independent of $Z_{i}$.
\end{theorem}

The reference distribution $Q$ can in fact be optimized, which would provide the tightest bound for a fixed learning algorithm. This bears certain resemblance to those used in \cite{dziugaite2017computing} which considers the computation of tight generalization bound using the PAC-Bayesian approach.

\begin{proof}
We start from (\ref{eqn:gen-ind1}), and consider each summand on the right-hand side
\begin{align}
&\Eb_{P_{W,Z_i}}\left[ L_{\xi}(W) - \ell(W, Z_i) \right]
=\Eb_{P_{Z_i}}\left[ \Eb_{P_{W|Z_i}}\left((L_{\xi}(W) - \ell(W, Z_i)\big{|} Z_i\right) \right]\notag\\
&\leq \Eb_{P_{Z_i}}\left[ \inf_{\lambda>0}\frac{\KL(P_{W|Z_i}||Q^i_W)+\Lambda_{F_i|Z_i,Q^i_W}(\lambda)}{\lambda}+\Eb_{Q^i_{W}}\left((L_{\xi}(W) - \ell(W, Z_i)\bigg{|} Z_i\right)\right]\notag\\
&=\Eb_{P_{Z_i}}\left[ \inf_{\lambda>0}\frac{\KL(P_{W|Z_i}||Q^i_W)+\Lambda_{F_i|Z_i,Q^i_W}(\lambda)}{\lambda}\right],
\end{align}
where the first equality is by the tower rule, the inequality is by (\ref{eqn:conditionalDV}), and the second equality is due to (\ref{eqn:bounda1a}). Summing over $i$ gives the bound stated in the theorem. 
\end{proof}

As will be shown in the next section, this bound is exactly tight for the quadratic Gaussian setting, and therefore, it can be viewed as a tight bound in the sense that it cannot be strictly improved in a uniform manner, either in terms of the constant or in the scaling. This bound can be loosened in several ways, which are stated in the following corollaries. 
\begin{corollary}\label{corr:first}
Let $F_i=L_{\xi}(W) - \ell(W,Z_i)$, then we have
\begin{align}
\gen(\xi, P_{W|Z_{[n]}}) 
&\leq  \frac{1}{n}\sum_{i=1}^n\inf_{\lambda > 0}\Eb\left[\frac{\KL(P_{W|Z_i}\|Q^i_{W})+\Lambda_{F_i,Q^i_{W}}(\lambda)}{\lambda}  \right]\notag\\
&\leq \inf_{\lambda > 0} \left[\frac{1}{n}\sum_{i=1}^n \Eb\left[\frac{\KL(P_{W|Z_i}\|Q^i_{W})+\Lambda_{F_i,Q^i_{W}}(\lambda)}{\lambda}  \right]\right],
\end{align}
for any $Q^i_{W,Z_i}=Q^i_WP_{Z_i}$, $i=1,2,\ldots,n$.
\end{corollary}
The first inequality is obtained by exchanging expectation and infimum, and the second is obtained by exchanging summation and infimum.
\begin{corollary}\label{corr:second}
Let $F_i=L_{\xi}(W) - \ell(W,Z_i)$, then we have
\begin{align}
\gen(\xi, P_{W|Z_{[n]}})
&\leq \Eb \inf_{\lambda > 0} \left[\frac{1}{n}\sum_{i=1}^n \left[\frac{\KL(P_{W|Z_i}\|Q^i_{W})+\Lambda_{F_i,Q^i_{W}}(\lambda)}{\lambda}  \right]\right]\notag\\
&\leq \inf_{\lambda > 0} \left[\frac{1}{n}\sum_{i=1}^n \Eb\left[\frac{\KL(P_{W|Z_i}\|Q^i_{W})+\Lambda_{F_i,Q^i_{W}}(\lambda)}{\lambda}  \right]\right]
\end{align}
for any $Q^i_{W,Z_i}=Q^i_WP_{Z_i}$, $i=1,2,\ldots,n$.
\end{corollary}
The first inequality is obtained by exchanging expectation and summation, and the second by exchanging infimum and expectation. The second bounds in Corollaries \ref{corr:first} and \ref{corr:second} are the same, while the first bounds are not directly comparable.

Notice that when $Q^i_{W,Z_i}=P_W\otimes P_{Z_i}$, i.e., the product of the marginals of $P_{W,{Z_i}}$, we have $\Eb[\KL(P_{W|Z_i}\|Q^i_{W})]=I(W;Z_i)$. This leads to the following corollary. 
\begin{corollary}\label{corr:third}
Let $F_i=L_{\xi}(W) - \ell(W,Z_i)$, then we have
\begin{align}
\gen(\xi, P_{W|Z_{[n]}})&\leq \frac{1}{n}\sum_{i=1}^n \inf_{\lambda > 0} \left[\frac{I(W;Z_i)+\Eb\Lambda_{F_i,P_{W}}(\lambda)}{\lambda}  \right]\notag\\
&\leq \frac{1}{n}\sum_{i=1}^n \inf_{\lambda > 0} \left[\frac{I(W;Z_i)+\Lambda_{F_i,P_{W}P_{Z_i}}(\lambda)}{\lambda}  \right],\label{eqn:MIa}\notag\\
&=\frac{1}{n}\sum_{i=1}^n \Lambda^{*-1}_{F_i,P_{W}P_{Z_i}}\left(I(W;Z_i)\right)
\end{align}
where the second inequality is due to the concavity of the $\ln(\cdot)$ function. 
\end{corollary}

By exchanging the infimum and the summation, we straightforwardly obtain further that
\begin{align}
\gen(\xi, P_{W|Z_{[n]}})
&\leq \inf_{\lambda > 0} \left[\frac{1}{n}\sum_{i=1}^n \left[\frac{I(W;Z_i)+\Eb\Lambda_{F_i,P_{W}}(\lambda)}{\lambda}  \right]\right]\notag\\
&\leq \inf_{\lambda > 0} \left[\frac{1}{n}\sum_{i=1}^n \left[\frac{I(W;Z_i)+\Lambda_{F_i,P_{W}P_{Z_i}}(\lambda)}{\lambda}  \right]\right]. \label{eqn:MIb}
\end{align}

The second bound in (\ref{eqn:MIa}) is in fact quite similar to the main theorem in \cite{bu2020tightening}. However, there is a major difference even when we assume the reference distribution $Q$ is the same as the product of the marginals in $P$: the function $F$ we choose to bound is different. % We shall return to this point in the next section. 

When the function $F$ is conditional $\sigma_{Z_i}$-sub-Gaussian with respect to the distribution $Q_{W}$, we have as a consequence $\Lambda_{F_i,Q^i_{W}}(\lambda)\leq \sigma_{Q_{Z_i}}^2\lambda^2$. 
The following corollary is then immediate.
\begin{corollary}
Let $F_i=L_{\xi}(W) - \ell(W,Z_i)$. If $F_i$ is conditional $\sigma_{Q_{z_i}}$-sub-Gaussian for each $Z_i=z_i$ with respective to $Q^i_W$ then
\begin{align}
\gen(\xi, P_{W|Z_{[n]}})&\leq \frac{1}{n}\sum_{i=1}^n \Eb \sqrt{\KL(P_{W|Z_i}\|Q^i_W)\sigma^2_{Q_{Z_i}}}\notag\\
&\leq \frac{1}{n}\sum_{i=1}^n \sqrt{\Eb\left[\KL(P_{W|Z_i}\|Q^i_W)\sigma^2_{Q_{Z_i}}\right]}.
\end{align}
for any $Q^i_{W}$ such that $W$ is independent of $Z_{i}$ for $i=1,2,\ldots,n$.
\end{corollary}

\section{The Canonical Quadratic Gaussian Setting Revisited}
\label{sec:Gaussian}

With the new generalization error bounds derived in the previous section, we are now ready to revisit the canonical quadratic Gaussian (location) setting.

\subsection{Exactly Tight Bounds for the Quadratic Gaussian Setting}
The expected generalization error of interest in the quadratic Gaussian setting is
\begin{align}
&\gen(\xi, P_{{W}|Z_{[n]}})  = \Eb\left[ \frac{1}{n} \sum_{i=1}^n \Eb\left[  \sigma^2 + \mu^2 -Z_i^2  + 2(Z_i - \mu) {W} | Z_i \right] \right].
\end{align}
For any fixed $i$, define 
\begin{align}
F_i=f_{Z_i}(W) \triangleq \sigma^2 + \mu^2 -Z_i^2 + 2(Z_i - \mu) W.\label{eqn:Fi}
\end{align}
Note the conditional distribution 
\begin{align}
{W}| Z_i \overset{P}{\sim} \mathcal{N}\left(\mu + \alpha_i (Z_i-\mu), \sum_{j\neq i} \alpha_j^2 \sigma^2 +\sigma_N^2\right).
\end{align}
We will choose the reference distribution $Q^i_{W}$ as
\begin{align}
W \overset{Q^i_W}{\sim} \mathcal{N}\left(\mu, \sum_{j\neq i} \alpha_j^2 \sigma^2 +\sigma_N^2\right), \label{eqn:QW}
\end{align}
which is indeed independent of $Z_i$. 

\vspace{0.2cm}
\noindent\textbf{Remark. } In the reference distribution $Q^i_{W,Z_i}$, $W$ and $Z_i$ are independent, and the marginal distribution $Q^i_W$ is not the same as that marginalized from $P_{W,Z_{[n]}}$. More specifically, the latter is in fact
$$P_W\sim \mathcal{N}\left(\mu, \sum_{i=1}^n \alpha_i^2\sigma^2+\sigma_N^2 \right),$$
which can be compared with (\ref{eqn:QW}). 

\vspace{0.2cm}

With these conditional distributions, we can derive that (see the appendix)
\begin{align}
&\KL(P_{{W}|Z_i} || Q^i_{W|Z_i})=\KL(P_{{W}|Z_i} || Q^i_{W}) = \alpha_i^2(Z_i - \mu)^2 \frac{1}{2\sum_{j\neq i} \alpha_j^2 \sigma^2 +2\sigma_N^2};\notag\\
&\Lambda_{F_i, Q^i_{W|Z_i}}(\lambda)=\Lambda_{F_i, Q^i_{W}}(\lambda)= 2\lambda^2 (Z_i - \mu)^2 \left(\sum_{j\neq i} \alpha_j^2 \sigma^2 +\sigma_N^2\right)  .
\end{align}
Therefore 
\begin{align}
&\Eb[\KL(P_{{W}|Z_i} || Q^i_{W})] = \alpha_i^2\sigma^2 \frac{1}{2\sum_{j\neq i} \alpha_j^2 \sigma^2 +2\sigma_N^2};\notag\\
&\Eb[\Lambda_{F_i, Q^i_{W}}(\lambda)] = 2  \lambda^2\sigma^2 \left(\sum_{j\neq i} \alpha_j^2 \sigma^2 +\sigma_N^2\right) .
\end{align}
Applying the first bound in Corollary \ref{corr:first}, we obtain
\begin{align}
\gen(\xi, P_{{W}|Z_{[n]}})
&\leq  \frac{1}{n}\sum_{i=1}^n \inf_{\lambda > 0}\Eb\left[\frac{\KL(P_{W|Z_i}\|Q^i_{W})+\Lambda_{F_i,Q^i_{W}}(\lambda)}{\lambda}  \right]\notag\\
&=\frac{1}{n}\sum_{i=1}^n \inf_{\lambda > 0} \left[\frac{\Eb[\KL(P_{W|Z_i}\|Q^i_{W})]+\Eb[\Lambda_{F_i,Q^i_{W}}(\lambda)]}{\lambda}  \right]\notag\\
&=\frac{2\sigma^2}{n},
\end{align}
where the last equality is by choosing the minimizer $\lambda_i^*$ as
\begin{align}
\lambda_i^*=\frac{\alpha_i}{2\sum_{j\neq i} \alpha_j^2 \sigma^2 +2\sigma_N^2}. \label{eqn:lambdastar}
\end{align}
Therefore, the first bound in Corollary \ref{corr:first} leads to a tight generalization error bound for this setting.

\vspace{0.2cm}
\noindent\textbf{Remark. } Recall the equality condition in (\ref{eqn:equalityconditiona}). With the given $P_{W|Z_i}$ and $Q^i_W$, we have that
\begin{align}
\ln\frac{dP}{dQ}=\frac{2\alpha_i(Z_i-\mu)W-\alpha_i^2(Z_i-\mu)^2-2\mu\alpha_i(Z_i-\mu)}{2\sum_{j\neq i} \alpha_j^2 \sigma^2 +2\sigma_N^2}.
\end{align}
With (\ref{eqn:Fi}), it is seen that the condition given in  (\ref{eqn:equalityconditiona}) is indeed satisfied, with
\begin{align}
\lambda=\frac{\alpha_i}{2\sum_{j\neq i} \alpha_j^2 \sigma^2 +2\sigma_N^2},\quad b=-\frac{\alpha_i^2(Z_i-\mu)^2+2\mu\alpha_i(Z_i-\mu)}{2\sum_{j\neq i} \alpha_j^2 \sigma^2 +2\sigma_N^2}.
\end{align}
This choice of $\lambda$ is in fact exactly the optimizing solution in (\ref{eqn:lambdastar}). Conversely, the distribution $Q$ we chose can be viewed as obtained through the condition (\ref{eqn:equalityconditiona}) (or equivalently $\dm Q\propto \exp(- \lambda f) \dm P$), with the parameter $\lambda$ chosen to maintain the independence between $W$ and $Z_i$ as required in Theorem \ref{theorem:main}. %If the reference distribution $Q^i_{W}$ can be chosen to make (\ref{eqn:equalitycondition}) hold, then the induced generalization error bound can be exactly tight unless subsequent bounding steps introduce some slackness. 
\vspace{0.2cm}

In contrast to the tight bound derived from the first bound in Corollary \ref{corr:first}, the second bound in Corollary \ref{corr:first} and the first bound in Corollary \ref{corr:second} are not tight for general assignments of $\alpha_i$'s, due to the fact that the optimal $\lambda^*_i$ is index-dependent. In the extreme case, consider setting $\alpha_1=1$ and $\alpha_i=0$ for $i=2,3,\ldots,n$. Then the second bound in Corollary \ref{corr:first} gives
\begin{align}
\gen(\xi, P_{{W}|Z_{[n]}})
&= \frac{1}{n}\inf_{\lambda > 0} \left[\frac{ \frac{\sigma^2}{2\sigma_N^2}+2(n-1) \sigma^2 \left(\sigma^2 +\sigma_N^2\right)\lambda^2+2\sigma^2\sigma_N^2\lambda^2}{\lambda}  \right]\notag\\
&= \frac{2\sigma^2}{n}\sqrt{\frac{2(n-1)\left(\sigma^2 +\sigma_N^2\right)+\sigma_N^2}{2\sigma_N^2}},
\end{align}
which is of order $\mathcal{O}(1/\sqrt{n})$. However, when $\alpha_i=1/n$, this dependence disappears and the loosened bounds also become tight. Indeed, consider the second bound  in Corollary \ref{corr:first} for this case, we have
\begin{align}
\gen(\xi, P_{{W}|Z_{[n]}})
&= \frac{1}{n}\inf_{\lambda > 0} \left[\frac{\sum_{i=1}^n(\Eb[\KL(P_{W|Z_i}\|Q^i_{W})]+\Eb[\Lambda_{F_i,Q^i_{W}}(\lambda)])}{\lambda}  \right]\notag\\
&= \frac{2\sigma^2}{n},
\end{align}
where the last step is obtained by choosing
\begin{align}
\lambda^*=\frac{\alpha_i}{2\sum_{j\neq i} \alpha_j^2 \sigma^2 +2\sigma_N^2}=\frac{n}{2(n-1) \sigma^2 +2n\sigma_N^2}.
\end{align}

\vspace{0.2cm}
\noindent\textbf{Remark. } The additive noise $N$ in the algorithm $W=\sum_{i=1}^n\alpha_i Z_i+N$ makes it a randomized algorithm, but it does not cause any essential difference in our bound. We included the noise here mostly to enlarge the set of problems that the proposed generalization error bound is tight. In other words, the proposed bound is not only tight for one particular algorithm of $\alpha_i=1/n$ and $\sigma_N^2=0$, but also a class of algorithms with different $\alpha_i$'s and $\sigma_N^2$. %Recall that the bound \cite{xu2017information} will be vacuous when $\sigma^2_N=0$, i.e., when the algorithm is deterministic, however, our proposed bound is derived along a similar line of individual-sample-based bound as that in \cite{bu2020tightening}, and therefore, it is not vacuous largely for the same reason as the individual-sample-based bound. Nevertheless, the tightness of the proposed bound is due to the unique features outlined previously.
\vspace{0.2cm}

\subsection{Looseness of Mutual Information Based Bounds}

One remaining question in the quadratic Gaussian setting is whether we can obtain tight or asymptotically tight generalization error bounds using mutual-information-based bounds. 
To understand this issue, we consider the bounds in Corollary \ref{corr:third} assuming the coefficients $\alpha_i=1/n$ for $i=1,2,\ldots,n$. Note that in this case, the choice of the reference distribution $Q^i_W$ is fixed as the marginal of $P_W$. 

The various terms we need when applying Corollary \ref{corr:third} in this setting can be shown to be (see the appendix)
\begin{align}
&I(W;Z_i)=\frac{1}{2}\log\frac{n}{n-1}\notag\\
&\Eb\Lambda_{F_i,Q^i_{W}}(\lambda)=\frac{2\sigma^4(n-1)}{n^2}\lambda^2\notag\\
&\Lambda_{F_i,Q^i_{W,Z_i}}(\lambda)=\lambda \sigma^2-\frac{1}{2}\log\left[1-2\left(\frac{2\lambda^2\sigma^4}{n}-\lambda \sigma^2\right)\right] \notag.
\end{align}

With these quantities, it follows that the first bound in Corollary \ref{corr:third} is
\begin{align}
&\gen(\xi, P_{{W}|Z_{[n]}})\leq \frac{2\sigma^2}{n}\sqrt{\left(\log\frac{n}{n-1}\right)\left(n-1\right)}.
\end{align}
The bound is of order $\mathcal{O}(1/n)$; in fact, it is asymptotically optimal in the sense that it approaches $\frac{2\sigma^2}{n}$. Therefore, the first mutual-information-based bound in Corollary \ref{corr:third} does not lose the tightness in a significant manner compared to the KL-based bound of those in Corollaries \ref{corr:first} and \ref{corr:second}.

The second bound in Corollary \ref{corr:third} has the form
\begin{align}
%&\gen(\xi, P_{{W}|Z_{[n]}})\leq  \sigma^2+\inf_{\lambda>0} \left[\frac{1}{2\lambda}\log\frac{n}{n-1}\right.\notag\\
%&\qquad\qquad\left.-\frac{1}{2\lambda}\log\left[1-2\left(\frac{2\lambda^2\sigma^4}{n}-\lambda \sigma^2\right)\right]\right],
&\gen(\xi, P_{{W}|Z_{[n]}})\leq  \sigma^2+\inf_{\lambda>0} \left[\frac{1}{2\lambda}\log\frac{n}{n-1}
-\frac{1}{2\lambda}\log\left[1-2\left(\frac{2\lambda^2\sigma^4}{n}-\lambda \sigma^2\right)\right]\right],
\end{align}
for any $\delta\in (0,1/2]$, and any $\epsilon>0$, by choosing $\lambda = 1/(2n^{\delta}\sigma^2)$, it can be seen that for sufficiently large $n$, we have $\gen(\xi, P_{{W}|Z_{[n]}})\leq (1+\epsilon) \frac{2\sigma^2}{n^{1-\delta}}$. Therefore, the bound can be also viewed as asymptotically optimal.

Similarly, we can apply the bounds in (\ref{eqn:MIb}). Since in this case, the optimal choice of $\lambda$ does not depend on the index-$i$, they are also asymptotically optimal. It should be noted that when the weight coefficients $\alpha_i$'s are not chosen to be uniform, then the optimal $\lambda$ becomes dependent on the index $i$, and the bounds in (\ref{eqn:MIb}) will be looser, in a similar manner as that for the KL-based bounds. 

{From the discussion on both the KL-based bound and the mutual-information-based bounds, it appears that the order-wise looseness of the existing bounds mainly stems from the choice of the function to apply the change of measure inequality, i.e., $\ell(W,Z_i)$ or $\ell(W,\tilde{Z})-\ell(W,Z_i)$. It is seen that the second quality is intuitively more centered, and therefore, the variance proxy is considerably lower than the former, assuming that they are both sub-Gaussian. In the canonical Gaussian setting, this difference is critical to make the information-theoretic bounds tight or asymptotically tight, and we expect the same effect will manifest in other problem settings, though without the ground truth and the statistical models, this conjecture is difficult to verify precisely. }

\section{Extension: The Vector Quadratic Gaussian Location Problem}
\label{sec:vector}

Let us consider the vector version of the quadratic Gaussian location problem. Let the data samples be $Z_1, Z_2, \ldots, Z_n \overset{i.i.d.}{\sim} \xi = \mathcal{N}(\mu, \Sigma)$, i.e., each $Z_i$ is a $d$-dimensional random Gaussian vector. The sample-average algorithm again chooses the following hypothesis $W = \sum_{i=1}^n\alpha_i Z_i+N$, where $\alpha_i$'s are nonnegative weights such that $\sum_{i = 1}^n \alpha_i =1$, and $N$ is a Gaussian noise vector $\sim \mathcal{N}(0,\sigma_N^2\mathbf{I})$. Instead of considering the standard mean squared error, let us consider a more general quadratic distortion measure  $\|x\|_A^2=x^T Ax$, based on a symmetric positive definite matrix $A$, for which we have 
\begin{align}
\gen(\xi, P_{W|Z_{[n]}})  &= \Eb\left[(\tilde{Z}-W)^TA(\tilde{Z}-W)- \frac{1}{n} \sum_{i=1}^n (Z_i - W)^TA(Z_i-W) \right]\notag\\
& = \frac{1}{n} \sum_{i=1}^n \left[\Tr(A(\Sigma+\mu\mu^T)) -\Eb\left(Z_i^T A Z_i  - 2(Z_i - \mu)^T AW \right)\right].
\end{align}
It can be shown that the generalization error of this setting is $\frac{2\Tr(A\Sigma)}{n}$.

One would expect that the result on the scalar setting could be generalized to this setting to obtain tight bounds, however, we shall illustrate the critical condition  (\ref{eqn:equalityconditiona}) is in fact rather stringent. To obtain tight results in this setting, one has to apply the bound in a different manner and the tightness is dependent on the decomposability of the loss function. 

\subsection{Generalization Error Bounds via Theorem \ref{theorem:main}}

Let us follow the footsteps of the scalar case, and define
\begin{align}
F_i=\Tr(A(\Sigma+\mu\mu^T))-\left(Z_i^T A Z_i  - 2(Z_i - \mu)^T AW \right). \label{eqn:Fi-vec}
\end{align}
The conditional distribution is
\begin{align}
{W}| Z_i \overset{P}{\sim} \mathcal{N}\left(\mu + \alpha_i (Z_i-\mu), \sum_{j\neq i} \alpha_j^2 \Sigma +\sigma_N^2\mathbf{I}\right).\label{eqn:PW-vec}
\end{align}
We will choose the reference distribution $Q^i_{W}$ as
\begin{align}
W \overset{Q^i_W}{\sim} \mathcal{N}\left(\mu, \sum_{j\neq i} \alpha_j^2 \Sigma +\sigma_N^2\mathbf{I}\right), \label{eqn:QW-vec}
\end{align}
which is independent of $Z_i$. 

With these conditional distributions, we can derive (see appendix) that
\begin{align}
&\KL(P_{{W}|Z_i} || Q^i_{W}) =\frac{\alpha_i^2}{2}\left[(Z_i-\mu)^T \left(\sum_{j\neq i} \alpha_j^2 \Sigma +\sigma_N^2\mathbf{I}\right)^{-1}(Z_i-\mu)\right]\label{eqn:KLvec}\\
&\Lambda_{F_i, Q^i_{W}}(\lambda) =2\lambda^2(Z_i-\mu)^TA\left(\sum_{j\neq i} \alpha_j^2 \Sigma +\sigma_N^2\mathbf{I}\right) A(Z_i-\mu).\label{eqn:Lambdavec}
\end{align}
Therefore 
\begin{align}
&\Eb[\KL(P_{{W}|Z_i} || Q^i_{W})] = \frac{\alpha_i^2}{2}\Tr\left[\left(\sum_{j\neq i}\alpha_j^2 \Sigma +\sigma_N^2\mathbf{I}\right)^{-1}\Sigma\right]\notag\\
&\Eb[\Lambda_{F_i, Q^i_{W}}(\lambda)] = 2\lambda^2 \Tr \left[A\left(\sum_{j\neq i}\alpha_j^2 \Sigma +\sigma_N^2\mathbf{I}\right) A\Sigma\right].
\end{align}

At this point, it is clear that the bounds can not be further simplified under general choices of $\alpha_i$'s, $A$, and $\sigma^2_N$. Next, we consider three special cases:
\begin{itemize}
\item $\sigma^2_N=0$ and $A=\mathbf{I}$: In this case, we have 
\begin{align}
\Eb[\KL(P_{{W}|Z_i} || Q^i_{W})] = \frac{d\alpha_i^2}{2\sum_{j\neq i}\alpha_j^2};\quad \Eb[\Lambda_{F_i, Q^i_{W}}(\lambda)] = 2\lambda^2\left(\sum_{j\neq i}\alpha_j^2\right)\Tr[\Sigma^2].
\end{align}
Applying the first bound in Corollary \ref{corr:first}, we obtain
\begin{align}
&\gen(\xi, P_{{W}|Z_{[n]}})\leq \frac{2}{n}\sqrt{d\Tr[\Sigma^2]}.
\end{align}
As a reference, the true generalization error in this setting is in fact $\frac{2}{n}\Tr[\Sigma]$, i.e., the bound is loose using this bounding approach.
\item $\sigma^2_N=0$ and $A=\Sigma^{-1}$:  In this case, we have 
\begin{align}
\Eb[\KL(P_{{W}|Z_i} || Q^i_{W})] = \frac{d\alpha_i^2}{2\sum_{j\neq i}\alpha_j^2};\quad \Eb[\Lambda_{F_i, Q^i_{W}}(\lambda)] = 2d\lambda^2\left(\sum_{j\neq i}\alpha_j^2\right).
\end{align}
Applying the first bound in Corollary \ref{corr:first}, we obtain
\begin{align}
&\gen(\xi, P_{{W}|Z_{[n]}})\leq \frac{2d}{n}.
\end{align}
For this case, the true generalization error is indeed fact $\frac{2d}{n}$, i.e., the bound is tight using this bounding approach. This setting is however a trivial setting, where the loss function essentially decomposes the vector into i.i.d. components.
\item $A=\mathbf{I}$, and $\Sigma = \sigma^2 \mathbf{I}$: In this case, we have 
\begin{align}
\Eb[\KL(P_{{W}|Z_i} || Q^i_{W})] = \frac{d\alpha_i^2\sigma^2}{2\sum_{j\neq i}\alpha_j^2\sigma^2+2\sigma_N^2};\quad \Eb[\Lambda_{F_i, Q^i_{W}}(\lambda)] = 2\lambda^2\left(\sum_{j\neq i}\alpha_j^2\sigma^2+\sigma^2_N\right)d\sigma^2.
\end{align}
Applying the first bound in Corollary \ref{corr:first}, we obtain
\begin{align}
&\gen(\xi, P_{{W}|Z_{[n]}})\leq \frac{2}{n}d\sigma^2.
\end{align}
The true generalization error in this setting is indeed the same, i.e., the bound is also tight for this special case. 
\end{itemize}

It is seen that in general the bounds derived from the proposed bounds given in  Theorem \ref{theorem:main} are not tight, but can yield tight bounds for certain special cases. 

\vspace{0.2cm}
\noindent\textbf{Remark. } Recall the equality condition in (\ref{eqn:equalityconditiona}). With the given $P_{W|Z_i}$ and $Q^i_W$, we have that
\begin{align}
\ln\frac{dP}{dQ}=&- (W-\mu-\alpha_i(Z_i-\mu))^T\left(\sum_{j\neq i}\alpha_j^2 \Sigma +\sigma_N^2\mathbf{I}\right)^{-1}(W-\mu-\alpha_i(Z_i-\mu))\notag\\
&\qquad +(W-\mu)^T\left(\sum_{j\neq i}\alpha_j^2 \Sigma +\sigma_N^2\mathbf{I}\right)^{-1}(W-\mu)\notag\\
=&2\alpha_i(Z_i-\mu)^T\left(\sum_{j\neq i}\alpha_j^2 \Sigma +\sigma_N^2\mathbf{I}\right)^{-1}(W-\mu).
\end{align}
With (\ref{eqn:Fi}), it is seen that the condition given in  (\ref{eqn:equalityconditiona}) can be satisfied when 
\begin{align}
\left(\sum_{j\neq i}\alpha_j^2 \Sigma +\sigma_N^2\mathbf{I}\right)^{-1}\propto A,
\end{align}
which indeed holds for the latter two cases discussed above. However, this relation does not hold under general $\Sigma$, $\sigma_N^2$, and $A$ choices, and bounds derived from Theorem \ref{theorem:main} will in general be loose.

\subsection{Generalization Error Bounds via Loss Function Decomposition}
Recall the loss function in general has the form $\ell:  \Wc  \times \Zc \rightarrow \Rb$. We call a loss function {\em decomposable} when there exist functions $\ell_j:  \Wc  \times \Zc \rightarrow \Rb$, $j=1,2,\ldots,d$, and functions $\phi_j: \Wc \rightarrow \Wc_j$, $j=1,2,\ldots,d$ such that
\begin{align}
\ell(w,z)=\sum_{j=1}^d \ell_j(\phi_j(w),z),
\end{align}
for any $(w,z)\in \Wc \times \Zc$. Clearly, the loss function we have adopted for the vector Gaussian location problem satisfies this condition with
\begin{align}
\ell_j(w,Z_i) & =\lambda_j (W-Z_i)^TU_jU^T_j(W-Z_i) \notag \\
& =(\sqrt{\lambda_j}U_j^TW-\sqrt{\lambda_j}U_j^TZ_i)^T(\sqrt{\lambda_j}U^T_jW-\sqrt{\lambda_j}U^T_jZ_i),
\end{align}
where $UDU^T$ is the eigenvalue decomposition of $A$, $\lambda_j$ is the $j$-th diagonal item of $D$, $U_j$ is the $j$-th column of $U$, and $\phi_j=\sqrt{\lambda_j}U^T_jW$. 

For decomposable loss functions, we have the following generalization of Theorem \ref{theorem:main}. 
\begin{theorem}
\label{theorem:main2}
Let $F_{i,j}=L_{j,\xi}(\phi_j(W)) - \ell_j(\phi_j(W),Z_i)$, then we have
\begin{align}
\gen(\xi, P_{W|Z_{[n]}})
&\leq\frac{1}{n}\sum_{i=1}^n \sum_{j=1}^d \Eb_{P_{Z_i}}\left[\inf_{\lambda > 0} \frac{\KL(P_{\phi_j(W)|Z_i}\|Q^i_{\phi_j(W)})+\Lambda_{F_{i,j}|Z_i,Q^i_{W}}(\lambda)}{\lambda}\right]\notag\\
&=  \frac{1}{n}\sum_{i=1}^n\sum_{j=1}^d \Eb_{P_{Z_i}}\left[\Lambda^{*-1}_{F_{i,j}|Z_i,Q^i_{W}}\left(\KL(P_{\phi_j(W)|Z_i}\|Q^i_{\phi_j(W)})\right)\right],
\end{align}
for $Q^i_{\phi_j(W),Z_i}=Q^i_{\phi_j(W)}P_{Z_i}$, $i=1,2,\ldots,n$, that is induced by any $Q^i_{W,Z_i}=Q^i_{W}P_{Z_i}$, $i=1,2,\ldots,n$, i.e., a distribution $Q^i$ where $W$ is independent of $Z_{i}$.
\end{theorem}

We omit its proof since it is almost identical to that of Theorem \ref{theorem:main}. It should be noted that the variational representation inequality is applied on the marginalized distribution $P_{\phi_j(W)|Z_i}$ and $Q^i_{\phi_j(W)}$, however since $Q^i_{\phi_j(W)}$ is induced by $Q^i_{W}$, we have $\Lambda_{F_{i,j}|Z_i,Q^i_{W}}(\lambda)=\Lambda_{F_{i,j}|Z_i,Q^i_{\phi(W)}}(\lambda)$. We provide the following corollary in order to tackle the vector Gaussian setting. A corollary similar to Corollary \ref{corr:second} can also be written, but it is omitted here for conciseness. 
\begin{corollary}\label{corr:first2}
Let $F_{i,j}=L_{j,\xi}(\phi_j(W)) - \ell_j(\phi_j(W),Z_i)$, then we have
\begin{align}
\gen(\xi, P_{W|Z_{[n]}})
&\leq  \frac{1}{n}\sum_{i=1}^n\sum_{j=1}^d\inf_{\lambda > 0}\Eb\left[\frac{\KL(P_{\phi_j(W)|Z_i}\|Q^i_{\phi_j(W)})+\Lambda_{F_{i,j},Q^i_{W}}(\lambda)}{\lambda}  \right]\notag\\
&\leq \inf_{\lambda > 0} \left[\frac{1}{n}\sum_{i=1}^n\sum_{j=1}^d \Eb\left[\frac{\KL(P_{\phi_j(W)|Z_i}\|Q^i_{\phi_j(W)})+\Lambda_{F_{i,j},Q^i_{W}}(\lambda)}{\lambda}  \right]\right],
\end{align}
for $Q^i_{\phi_j(W),Z_i}=Q^i_{\phi_j(W)}P_{Z_i}$, $i=1,2,\ldots,n$, that is induced by any $Q^i_{W,Z_i}=Q^i_{W}P_{Z_i}$, $i=1,2,\ldots,n.$
\end{corollary}

Equipped with the new bounds above, let us revisit the vector setting. This time, let us define
\begin{align}
F_{i,j}=\lambda_j\Tr(U_jU_j^T(\Sigma+\mu\mu^T))-\lambda_j\left(Z_i^T U_jU_j^T Z_i  - 2(Z_i - \mu)^T U_jU_j^TW \right).
\end{align}
The conditional distribution $P_{\phi_j(W)|Z_i}$ is given as 
\begin{align}
P_{\phi_j(W)|Z_i}=P_{\sqrt{\lambda_j}U_j^\top W | Z_i} = \mathcal{N}\left(\sqrt{\lambda_j}U_j^\top \mu + \alpha_i\sqrt{\lambda_j} U_j^\top (Z_i-\mu), \lambda_j\sum_{j\neq i} \alpha_j^2 U_j^\top \Sigma U_j +\lambda_j\sigma_N^2\right).
\end{align}
We will choose the reference distribution $Q^i_{\phi_j(W)}=Q^i_{\sqrt{\lambda_j}U_j^TW}$ as
\begin{align}
\sqrt{\lambda_j}U_j^\top W {\,\sim\,} \mathcal{N}\left(\sqrt{\lambda_j}U_j^\top \mu, \lambda_j\sum_{j\neq i} \alpha_j^2 U_j^\top \Sigma U_j +\lambda_j\sigma_N^2 \right), 
\end{align}
The divergence term $\KL(P_{\phi_j(W)|Z_i} || Q^i_{\phi_j(W)})$  is therefore
\begin{align}
\KL( P_{\sqrt{\lambda_j}U_j^\top W|Z_i} || Q^i_{\sqrt{\lambda_j}U_j^\top W} ) = \frac{\alpha_i^2 \Tr[U_jU_j^T (Z_i-\mu)(Z_i-\mu)^T] }{ 2\left(\sum_{j\neq i} \alpha_j^2 U_j^\top \Sigma U_j +\sigma_N^2\right) }.
\end{align}
By substituting $A=U_jU_J^T$ in (\ref{eqn:Lambdavec}), we can obtain that
\begin{align}
\Lambda_{F_{i,j}, Q^i_{W}}(\lambda) &=2\lambda^2_j\lambda^2(Z_i-\mu)^T U_jU_j^T\left(\sum_{j\neq i} \alpha_j^2 \Sigma +\sigma_N^2\mathbf{I}\right)  U_jU_j^T(Z_i-\mu)\notag\\
&=2\lambda_j^2\lambda^2\left(\sum_{j\neq i} \alpha_j^2 U_j^T\Sigma U_j +\sigma_N^2\right) \Tr[U_jU_j^T (Z_i-\mu)(Z_i-\mu)^T]
\end{align}
Therefore 
\begin{align}
\Eb[\KL(P_{\phi_j(W)|Z_i} || Q^i_{\phi(W)})] &= \frac{\alpha_i^2 \Tr[U_jU_j^T\Sigma] }{ 2\left(\sum_{j\neq i} \alpha_j^2 U_j^\top \Sigma U_j +\sigma_N^2\right) }\notag\\
\Eb[\Lambda_{F_{i,j}, Q^i_{W}}(\lambda)] &= 2\lambda_j^2\lambda^2\left(\sum_{j\neq i} \alpha_j^2 U_j^T\Sigma U_j +\sigma_N^2\right) \Tr[U_jU_j^T \Sigma].
\end{align}

Applying the first bound in Corollary \ref{corr:first2}, we obtain
\begin{align}
\gen(\xi, P_{{W}|Z_{[n]}})
&\leq  \frac{1}{n}\sum_{i=1}^n\sum_{j=1}^d \inf_{\lambda > 0}\Eb\left[\frac{\KL(P_{\phi_j(W)|Z_i} || Q^i_{\phi(W)})+\Lambda_{F_{i,j}, Q^i_{W}}(\lambda)}{\lambda}  \right]\notag\\
&=\frac{1}{n}\sum_{i=1}^n\sum_{j=1}^d 2\alpha_j\lambda_j\Tr[U_jU_j^T \Sigma]= \frac{2}{n}\sum_{j=1}^d\lambda_j\Tr\left[ U_jU_j^T \Sigma\right]\notag\\
&=\frac{2}{n}\Tr\left[\sum_{j=1}^d\lambda_j U_jU_j^T \Sigma\right]=\frac{2}{n}\Tr\left[A \Sigma\right],
\end{align}
which is indeed the true generalization error.

\vspace{0.2cm}
\noindent\textbf{Remark:} Though loss functions may not be decomposable in general, for the vector Gaussian location problem, decomposability can indeed be utilized to yield a tight information-theoretic generalization bound, as shown above. In a sense, decomposition allows us to utilize the probability distribution of a random variable after further processing, and by the data-processing inequality of KL divergence \cite{wu2019information}, such processing will reduce the KL divergence and potentially yield tighter bounds.  

\section{Conclusion}
\label{sec:conclusion}

We studied the information-theoretic generalization error bounds, and in particular, focused on the quadratic Gaussian setting. The proposed new bound is shown to be exactly tight for this setting. The most important change from the previous work appears to be the function that we choose to bound, however, the additional introduction of a reference distribution, and the conditional application of the change of measure inequality also contribute to the tightness of the bound. A generalized vector version of the problem is further studied, which inspired a new and refined generalization error bound that relies on the decomposability of the loss functions.

{
Though we have focused on the quadratic Gaussian setting exclusively in this work, the technique can be applied to the study of noisy and iterative algorithms such as stochastic gradient Langevin dynamics (SGLD), as previously studied in \cite{pensia2018generalization,rodriguez2021random, bu2020tightening,haghifam2020sharpened}. The key difference from the previous result is that due to the application of the change of measure inequality, our bound relies on the cumulant generating function of a different quantity, or a different sub-Gaussian variance proxy, that likely has a lower value, and therefore the resultant bound is also potentially tighter in that setting. However, due to the more complex statistical dependence induced by the algorithm, it is not clear whether this can drive order-wise gains, and we leave this to a future study.}

Gaussian models have had many successes in machine learning research, particularly in the context of Gaussian process \cite{williams2006gaussian} and the recent development of Gaussian diffusion models \cite{sohl2015deep,ho2020denoising,song2020score,dhariwal2021diffusion,negrea2019information}. Therefore, we believe studying the Gaussian settings in the context of machine learning is indeed well-motivated, and will lead to important engineering insights in the future.

\appendix
\section{Generalization Error for the Quadratic Gaussian Case}
We can write as follows to derive the exact generalization error for the canonical quadratic Gaussian setting without utilizing the information-theoretical bounds as follows: 
\begin{align}
&\gen(\xi, P_{W|Z_{[n]}})  = \Eb\left[(\tilde{Z}-W)^2- \frac{1}{n} \sum_{i=1}^n(Z_i - W)^2 \right]\notag\\
&\qquad = \frac{1}{n} \sum_{i=1}^n \Eb\left( \sigma^2 + \mu^2 -Z_i^2  + 2(Z_i - \mu) W \right)\notag\\
&\qquad = \frac{2}{n} \sum_{i=1}^n \Eb\left[\left(Z_i - \mu\right) \left(\sum_{j=1}^n \alpha_j Z_j+N\right) \right]\notag\\
%&\qquad = \frac{2}{n} \Eb\left[\sum_{i=1}^n (Z_i - \mu)\right) \left(\sum_{i=1}^n \alpha_i Z_i+N\right) \right]\notag\\
&\qquad = \frac{2}{n} \sum_{i=1}^n \Eb\left[(Z_i - \mu) \left(\sum_{j=1}^n \alpha_j (Z_j-\mu)+N\right)\right]\notag\\
%&\qquad = \frac{2}{n} \Eb\left[\sum_{i=1}^n (Z_i - \mu)\right) \left(\sum_{i=1}^n \alpha_i (Z_i-\mu)\right)\right]\notag\\
&\qquad = \frac{2}{n} \sum_{i=1}^n\sum_{j=1}^n\alpha_j \Eb\left[(Z_i - \mu) (Z_j-\mu)\right]\notag\\
&\qquad = \frac{2}{n} \sum_{i=1}^n\alpha_i \Eb\left[(Z_i - \mu)^2\right]=\frac{2\sigma^2}{n}.
\end{align}
This gives the exact generalization error for this setting. 

\section{Computing $\Lambda_{F_i,Q^i_{W|Z_i}}(\lambda)=\Lambda_{F_i,Q^i_{W}}(\lambda)$: The Scalar Gaussian Case}
First, notice that 
\begin{align}
\Eb_{Q^i_{W|Z_i}} [F_i] &=\Eb_{Q^i_{W|Z_i}} \left[(\sigma^2 + \mu^2 -Z_i^2) + 2(Z_i - \mu) W|Z_i\right]\notag\\
&=(\sigma^2 + \mu^2 -Z_i^2) + 2\mu(Z_i - \mu),
\end{align}
since under $Q^i{W|Z_i}$, $W$ and $Z_i$ are independent, and $W$ has mean $\mu$. 
Then we can write
\begin{align}
&\Eb_{Q^i_{W|Z_i}}\exp\left[\lambda(\sigma^2 + \mu^2 -Z_i^2) + 2\lambda(Z_i - \mu) W\right|Z_i]\notag\\
&=\exp[\lambda(\sigma^2 + \mu^2 -Z_i^2)]\Eb_{Q^i_{W|Z_i}}[\exp\left(2\lambda(Z_i - \mu) W\right)|Z_i]\notag\\
&=\exp[\lambda(\sigma^2 + \mu^2 -Z_i^2)] \exp[2\lambda\mu(Z_i - \mu)]\exp\left[2\lambda^2(Z_i - \mu)^2\left(\sum_{j\neq i} \alpha_j^2 \sigma^2 +\sigma_N^2\right)\right],
\end{align}
where the second equality is by the moment generating function of Gaussian random variable $W$ distributed according to $Q^i_{W}$. It follows then
\begin{align}
\Lambda_{F_i,Q^i_{W|Z_i}}(\lambda)=\ln \Eb_{Q^i_{W|Z_i}}[\exp(\lambda F_i)]-\lambda\Eb[F_i]=2\lambda^2(Z_i - \mu)^2\left(\sum_{j\neq i} \alpha_j^2 \sigma^2 +\sigma_N^2\right).
\end{align}

\section{Computing $\Lambda_{F_i,Q_{W,Z_i}}(\lambda)=\Lambda_{F_i,P_{W}P_{Z_i}}(\lambda)$: The Scalar Gaussian Case}
First, notice that $\Eb_{P_{W}P_{Z_i}} [F_i]=0$. Then 
\begin{align}
&\Eb_{P_{W}P_{Z_i}}\exp\left(\lambda(\sigma^2 + \mu^2 -Z_i^2) + 2\lambda(Z_i - \mu) W\right)\notag\\
&=\Eb\left[\Eb[\exp\left(\lambda(\sigma^2 + \mu^2 -Z_i^2) + 2\lambda(Z_i - \mu) W\right)|Z_i]\right]\notag\\
&=\Eb\left[\exp(\lambda(\sigma^2 + \mu^2 -Z_i^2 )) \cdot\exp(2\lambda(Z_i - \mu)\mu+2\lambda^2(Z_i - \mu)^2\sigma^2/n)|Z_i\right],
\end{align}
where the first equality is by the tower rule, and the second step is by using the moment generating function of the Gaussian random variable $W$. 
Rearranging the terms gives
\begin{align} 
&\Eb_{P_{W}P_{Z_i}}\exp\left(\lambda(\sigma^2 + \mu^2 -Z_i^2) + 2\lambda(Z_i - \mu) W\right)\notag\\
&=\exp(\lambda\sigma^2)\Eb\exp\left[\left(\frac{2\lambda^2\sigma^2}{n}-\lambda\right)(Z_i-\mu)^2\right]\notag\\
&=\exp(\lambda\sigma^2)\left(1-2\left(\frac{2\lambda^2\sigma^2}{n}-\lambda\right)\sigma^2\right)^{-1/2},
\end{align}
where the last equality is by the moment generating function of the $\chi^2$ random variable of degree one. 
Taking the logarithm on the right-hand side gives the expression for $\Lambda_{F_i,P_{W}P_{Z_i}}(\lambda)$.

\section{Computing $\Lambda_{F_i,Q^i_{W|Z_i}}(\lambda)=\Lambda_{F_i,Q^i_{W}}(\lambda)$: The Vector Gaussian Case}
Similar to the scalar case, notice that 
\begin{align}
\Eb_{Q^i_{W|Z_i}} [ F_i] &=\Eb_{Q^i_{W|Z_i}} \left[\Tr(A(\Sigma+\mu\mu^T))-\left(Z_i^T A Z_i  - 2(Z_i - \mu)^T AW \right)|Z_i\right]\notag\\
&=(\Tr(A(\Sigma+\mu\mu^T))-Z_i^T A Z_i) + 2(Z_i - \mu)^T A \mu.
\end{align}
We can then write the exponential term in $\Lambda_{F_i,Q^i_{W|Z_i}}(\lambda)$
\begin{align}
&\Eb_{Q^i_{W|Z_i}}\exp\left[\lambda\Tr(A(\Sigma+\mu\mu^T))-\lambda\left(Z_i^T A Z_i  - 2(Z_i - \mu)^T AW \right)|Z_i\right]\notag\\
&=\exp[\lambda\Tr(A(\Sigma+\mu\mu^T))-\lambda Z_i^T A Z_i]\Eb_{Q^i_{W|Z_i}}\left[\exp\left(2\lambda(Z_i - \mu)^T AW\right)|Z_i\right]\notag\\
&=\exp[\lambda\Tr(A(\Sigma+\mu\mu^T))-\lambda Z_i^T A Z_i] \exp\left[ 2\lambda(Z_i - \mu)^TA\mu \right]\notag\\
&\qquad\qquad \cdot\exp\left[2\lambda^2(Z_i-\mu)^TA\left(\sum_{j\neq i} (\alpha_j^2 \Sigma) +\sigma_N^2\mathbf{I}\right) A^T(Z_i-\mu)\right]
%&=\exp[\lambda(\sigma^2 + \mu^2 -Z_i^2)] \exp[2\lambda\mu(Z_i - \mu)]\exp\left[2\lambda^2(Z_i - \mu)^2\left(\sum_{j\neq i} \alpha_j^2 \sigma^2 +\sigma_N^2\right)\right],
\end{align}
where the second equality follows standard manipulation of Gaussian integration. It follows then
\begin{align}
\Lambda_{F_i,Q^i_{W|Z_i}}(\lambda)=\ln \Eb{Q^i_{W|Z_i}}[\exp(\lambda F_i)]-\lambda\Eb[F_i]=2\lambda^2(Z_i-\mu)^TA\left(\sum_{j\neq i} (\alpha_j^2 \Sigma) +\sigma_N^2\mathbf{I}\right) A^T(Z_i-\mu).
\end{align}

\bibliographystyle{IEEEtran}

\begin{thebibliography}{10}
\providecommand{\url}[1]{#1}
\csname url@samestyle\endcsname
\providecommand{\newblock}{\relax}
\providecommand{\bibinfo}[2]{#2}
\providecommand{\BIBentrySTDinterwordspacing}{\spaceskip=0pt\relax}
\providecommand{\BIBentryALTinterwordstretchfactor}{4}
\providecommand{\BIBentryALTinterwordspacing}{\spaceskip=\fontdimen2\font plus
\BIBentryALTinterwordstretchfactor\fontdimen3\font minus
  \fontdimen4\font\relax}
\providecommand{\BIBforeignlanguage}[2]{{%
\expandafter\ifx\csname l@#1\endcsname\relax
\typeout{** WARNING: IEEEtran.bst: No hyphenation pattern has been}%
\typeout{** loaded for the language `#1'. Using the pattern for}%
\typeout{** the default language instead.}%
\else
\language=\csname l@#1\endcsname
\fi
#2}}
\providecommand{\BIBdecl}{\relax}
\BIBdecl

\bibitem{russo2016controlling}
D.~Russo and J.~Zou, ``Controlling bias in adaptive data analysis using
  information theory,'' in \emph{Artificial Intelligence and Statistics}, 2016,
  pp. 1232--1240.

\bibitem{xu2017information}
A.~Xu and M.~Raginsky, ``Information-theoretic analysis of generalization
  capability of learning algorithms,'' in \emph{Advances in Neural Information
  Processing Systems}, 2017, pp. 2524--2533.

\bibitem{asadi2018chaining}
A.~Asadi, E.~Abbe, and S.~Verd{\'u}, ``Chaining mutual information and
  tightening generalization bounds,'' in \emph{Advances in Neural Information
  Processing Systems}, 2018, pp. 7234--7243.

\bibitem{pensia2018generalization}
A.~Pensia, V.~Jog, and P.-L. Loh, ``Generalization error bounds for noisy,
  iterative algorithms,'' in \emph{2018 IEEE International Symposium on
  Information Theory (ISIT)}, Jun. 2018, pp. 546--550.

\bibitem{issa2019strengthened}
I.~Issa, A.~R. Esposito, and M.~Gastpar, ``Strengthened information-theoretic
  bounds on the generalization error,'' in \emph{2019 IEEE International
  Symposium on Information Theory (ISIT)}, Jul. 2019, pp. 582--586.

\bibitem{negrea2019information}
J.~Negrea, M.~Haghifam, G.~K. Dziugaite, A.~Khisti, and D.~M. Roy,
  ``Information-theoretic generalization bounds for {SGLD} via data-dependent
  estimates,'' in \emph{Advances in Neural Information Processing Systems},
  2019, pp. 11\,015--11\,025.

\bibitem{jose2021information}
S.~T. Jose and O.~Simeone, ``Information-theoretic generalization bounds for
  meta-learning and applications,'' \emph{Entropy}, vol.~23, no.~1, p. 126,
  2021.

\bibitem{wu2020information}
X.~Wu, J.~H. Manton, U.~Aickelin, and J.~Zhu, ``Information-theoretic analysis
  for transfer learning,'' in \emph{2020 IEEE International Symposium on
  Information Theory (ISIT)}, Jun. 2020, pp. 2819--2824.

\bibitem{bu2020tightening}
Y.~Bu, S.~Zou, and V.~V. Veeravalli, ``Tightening mutual information based
  bounds on generalization error,'' \emph{IEEE Journal on Selected Areas in
  Information Theory}, vol.~1, no.~1, pp. 121--130, 2020.

\bibitem{steinke2020reasoning}
T.~Steinke and L.~Zakynthinou, ``Reasoning about generalization via conditional
  mutual information,'' in \emph{Conference on Learning Theory}.\hskip 1em plus
  0.5em minus 0.4em\relax PMLR, 2020, pp. 3437--3452.

\bibitem{haghifam2020sharpened}
M.~Haghifam, J.~Negrea, A.~Khisti, D.~M. Roy, and G.~K. Dziugaite, ``Sharpened
  generalization bounds based on conditional mutual information and an
  application to noisy, iterative algorithms,'' in \emph{Advances in Neural
  Information Processing Systems}, vol.~33, 2020, pp. 9925--9935.

\bibitem{hellstrom2020generalization}
F.~Hellstr{\"o}m and G.~Durisi, ``Generalization bounds via information density
  and conditional information density,'' \emph{IEEE Journal on Selected Areas
  in Information Theory}, vol.~1, no.~3, pp. 824--839, 2020.

\bibitem{wu2022fast}
X.~Wu, J.~H. Manton, U.~Aickelin, and J.~Zhu, ``Fast rate generalization error
  bounds: Variations on a theme,'' \emph{arXiv preprint arXiv:2205.03131},
  2022.

\bibitem{zhou2022individually}
R.~Zhou, C.~Tian, and T.~Liu, ``Individually conditional individual mutual
  information bound on generalization error,'' \emph{IEEE Transactions on
  Information Theory}, vol.~68, no.~5, pp. 3304--3316, 2022.

\bibitem{rodriguez2021random}
B.~Rodr{\'\i}guez-G{\'a}lvez, G.~Bassi, R.~Thobaben, and M.~Skoglund, ``On
  random subset generalization error bounds and the stochastic gradient
  {Langevin} dynamics algorithm,'' in \emph{Proc. 2020 IEEE Information Theory
  Workshop (ITW)}, Apr. 2021, pp. 1--5.

\bibitem{zhou2022stochastic}
R.~Zhou, C.~Tian, and T.~Liu, ``Stochastic chaining and strengthened
  information-theoretic generalization bounds,'' in \emph{2019 IEEE
  International Symposium on Information Theory (ISIT)}, 2022, pp. 690--695.

\bibitem{aminian2021exact}
G.~Aminian, Y.~Bu, L.~Toni, M.~Rodrigues, and G.~Wornell, ``An exact
  characterization of the generalization error for the {Gibbs} algorithm,'' in
  \emph{Advances in Neural Information Processing Systems}, vol.~34, 2021, pp.
  8106--8118.

\bibitem{barnes2022improved}
L.~P. Barnes, A.~Dytso, and H.~V. Poor, ``Improved information-theoretic
  generalization bounds for distributed, federated, and iterative learning,''
  \emph{Entropy}, vol.~24, no.~9, p. 1178, 2022.

\bibitem{aminian2022tighter}
G.~Aminian, Y.~Bu, G.~Wornell, and M.~Rodrigues, ``Tighter expected
  generalization error bounds via convexity of information measures,''
  \emph{arXiv preprint arXiv:2202.12150}, 2022.

\bibitem{haghifam2022understanding}
M.~Haghifam, S.~Moran, D.~M. Roy, and G.~K. Dziugiate, ``Understanding
  generalization via leave-one-out conditional mutual information,'' in
  \emph{2022 IEEE International Symposium on Information Theory (ISIT)}, Jul.
  2022, pp. 2487--2492.

\bibitem{hafez2020conditioning}
H.~Hafez-Kolahi, Z.~Golgooni, S.~Kasaei, and M.~Soleymani, ``Conditioning and
  processing: Techniques to improve information-theoretic generalization
  bounds,'' \emph{Advances in Neural Information Processing Systems}, vol.~33,
  2020.

\bibitem{hellstrom2022evaluated}
F.~Hellstr{\"o}m and G.~Durisi, ``Evaluated {CMI} bounds for meta learning:
  Tightness and expressiveness,'' \emph{arXiv preprint arXiv:2210.06511}, 2022.

\bibitem{haghifam2023limitations}
M.~Haghifam, B.~Rodr{\'\i}guez-G{\'a}lvez, R.~Thobaben, M.~Skoglund, D.~M. Roy,
  and G.~K. Dziugaite, ``Limitations of information-theoretic generalization
  bounds for gradient descent methods in stochastic convex optimization,'' in
  \emph{International Conference on Algorithmic Learning Theory}.\hskip 1em
  plus 0.5em minus 0.4em\relax PMLR, 2023, pp. 663--706.

\bibitem{wang2023generalization}
H.~Wang, R.~Gao, and F.~P. Calmon, ``Generalization bounds for noisy iterative
  algorithms using properties of additive noise channels,'' \emph{J. Mach.
  Learn. Res.}, vol.~24, pp. 26--1, 2023.

\bibitem{wang2023tighter}
Z.~Wang and Y.~Mao, ``Tighter information-theoretic generalization bounds from
  supersamples,'' \emph{arXiv preprint arXiv:2302.02432}, 2023.

\bibitem{wang2021generalization}
------, ``On the generalization of models trained with {SGD}:
  Information-theoretic bounds and implications,'' \emph{arXiv preprint
  arXiv:2110.03128}, 2021.

\bibitem{shalev2014understanding}
S.~Shalev-Shwartz and S.~Ben-David, \emph{Understanding machine learning: From
  theory to algorithms}.\hskip 1em plus 0.5em minus 0.4em\relax Cambridge
  university press, 2014.

\bibitem{CoverThomas}
T.~M. Cover and J.~A. Thomas, \emph{Elements of Information Theory},
  1st~ed.\hskip 1em plus 0.5em minus 0.4em\relax New York: Wiley, 1991.

\bibitem{Shannon:48}
C.~E. Shannon, ``A mathematical theory of communication,'' \emph{Bell System
  Technical Journal}, vol.~27, pp. 379--423, 623--656, Jul. Oct. 1948.

\bibitem{tse2005fundamentals}
D.~Tse and P.~Viswanath, \emph{Fundamentals of wireless communication}.\hskip
  1em plus 0.5em minus 0.4em\relax Cambridge university press, 2005.

\bibitem{berger2003rate}
T.~Berger, ``Rate-distortion theory,'' \emph{Wiley Encyclopedia of
  Telecommunications}, 2003.

\bibitem{gray1989source}
R.~M. Gray, \emph{Source coding theory}.\hskip 1em plus 0.5em minus 0.4em\relax
  Springer Science \& Business Media, 1989, vol.~83.

\bibitem{hellstrom2023generalization}
F.~Hellstr{\"o}m, G.~Durisi, B.~Guedj, and M.~Raginsky, ``Generalization
  bounds: Perspectives from information theory and {PAC-Bayes},'' \emph{arXiv
  preprint arXiv:2309.04381}, 2023.

\bibitem{dziugaite2017computing}
G.~K. Dziugaite and D.~M. Roy, ``Computing nonvacuous generalization bounds for
  deep (stochastic) neural networks with many more parameters than training
  data,'' in \emph{Proceedings of the Conference on Uncertainty in Artificial
  Intelligence}, 2017.

\bibitem{wu2019information}
Y.~Wu, ``Information-theoretic methods for high-dimensional statistics,'' 2019.

\bibitem{williams2006gaussian}
C.~K. Williams and C.~E. Rasmussen, \emph{Gaussian processes for machine
  learning}.\hskip 1em plus 0.5em minus 0.4em\relax MIT press Cambridge, MA,
  2006, vol.~2, no.~3.

\bibitem{sohl2015deep}
J.~Sohl-Dickstein, E.~Weiss, N.~Maheswaranathan, and S.~Ganguli, ``Deep
  unsupervised learning using nonequilibrium thermodynamics,'' in
  \emph{International Conference on Machine Learning}.\hskip 1em plus 0.5em
  minus 0.4em\relax PMLR, 2015, pp. 2256--2265.

\bibitem{ho2020denoising}
J.~Ho, A.~Jain, and P.~Abbeel, ``Denoising diffusion probabilistic models,''
  \emph{Advances in Neural Information Processing Systems}, vol.~33, pp.
  6840--6851, 2020.

\bibitem{song2020score}
Y.~Song, J.~Sohl-Dickstein, D.~P. Kingma, A.~Kumar, S.~Ermon, and B.~Poole,
  ``Score-based generative modeling through stochastic differential
  equations,'' \emph{arXiv preprint arXiv:2011.13456}, 2020.

\bibitem{dhariwal2021diffusion}
P.~Dhariwal and A.~Nichol, ``Diffusion models beat {GANs} on image synthesis,''
  \emph{Advances in Neural Information Processing Systems}, vol.~34, pp.
  8780--8794, 2021.

\end{thebibliography}
 \newcommand{\noop}[1]{}

\end{document}